\documentclass[letterpaper, 10 pt, conference]{ieeeconf}  

\IEEEoverridecommandlockouts                              

\usepackage{graphics} 
\usepackage{epstopdf}
\usepackage{epsfig} 
\usepackage{amsmath} 
\usepackage{amssymb}  
\usepackage{midfloat}
\usepackage{soul}

\usepackage{color}

\newcommand{\bi}{\textcolor{blue}}

\newcommand{\forceindent}{\leavevmode{\parindent=1em\indent}}

\title{\LARGE \bf
Detection and Isolation of Small Faults in Lithium-Ion Batteries \\ via the Asymptotic Local Approach*
}

\author{Luis D.\ Couto$^{1,2}$, Jorn M. Reniers$^{1}$, David A. Howey$^{1,3}$ and Michel Kinnaert$^{2}$
\thanks{*L.D.\ Couto would like to thank the Wiener-Anspach Foundation for its financial support. This work was supported by the Fond de la Recherche Scientifique - FNRS under grant n$^\circ$T.0142.20}
\thanks{$^{1}$L.D.\ Couto ({\tt\small luis.coutomendonca@eng.ox.ac.uk}), J.M.\ Reniers ({\tt\small jorn.reniers@eng.ox.ac.uk}) and D.A.\ Howey ({\tt\small david.howey@eng.ox.ac.uk}) are with the Department of Engineering Science, University of Oxford, Oxford OX1 3PJ, UK.}%
\thanks{$^{2}$L.D. Couto and M. Kinnaert ({\tt\small michel.kinnaert@ulb.ac.be}) are with with the School of Engineering of the Universit\'e Libre de Bruxelles, B-1050 Brussels, Belgium.}
\thanks{$^{3}$D.A.\ Howey is also with the Faraday Institution, Harwell Campus, Didcot, OX11 0RA, UK.}%
}

\begin{document}

\maketitle
\thispagestyle{empty}
\pagestyle{empty}

\begin{abstract}
This contribution presents a diagnosis scheme for batteries to detect and isolate internal faults in the form of small parameter changes. This scheme is based on an electrochemical reduced-order model of the battery, which allows the inclusion of physically meaningful faults that might affect the battery performance. 
The sensitivity properties of the model are analyzed. 
The model is then used to compute residuals based on an unscented Kalman filter. Primary residuals and a limiting covariance matrix are obtained thanks to the local approach, allowing for fault detection and isolation by $\chi^2$ statistical tests. 
Results show that faults resulting in limited 0.15\% capacity and 0.004\% power fade can be effectively detected by the local approach. The algorithm is also able to correctly isolate faults related with sensitive parameters, whereas parameters with low sensitivity or linearly correlated are more difficult to precise.
\end{abstract}

\section{INTRODUCTION}

Lithium-ion (Li-ion) batteries are all around us, from  mobile phones to electric vehicles. This ubiquity comes from the fact that these energy storage devices have a high energy and power density, compared with other types of batteries \cite{Chaturvedi-2010}. However, this improved performance comes at the cost of safety issues, such as possible short circuit or thermal runaway \cite{Onori-2020}, if Li-ion batteries are mistreated. A common way to prevent possible safety hazards in Li-ion batteries is to oversize them, which results in more bulky and expensive batteries whose operation is overly conservative \cite{Chaturvedi-2010}. 
An alternative way to guarantee safe operation without incurring unnecessary cost is to monitor battery condition through model-based approaches. Among other tasks, such a monitoring system is in charge of performing battery diagnostics, i.e. detection and isolation of possible faults affecting battery operation. In this context, special emphasis is placed on early warnings in order to prevent failure and predict maintenance.

Efforts in fault diagnostics for Li-ion batteries have grown in recent years \cite{Onori-2020}. 
They fall into two main categories depending on the diagnostic method used, namely data-driven and model-based methods \cite{Onori-2020}. 
Data-driven methods provide high-order nonlinear approximations that are versatile to accommodate the estimated faults in relation to the measured data, but they suffer from requiring large and rich fault datasets for model training and they are disconnected from physical principles. 
These pitfalls are overcome by model-based methods. This second category of methods aims at detecting and isolating changes in greybox model parameters, which are linked to the state-of-health of the battery. 
Within this category, a first and the most popular approach is to estimate both parameters \& states using parameter estimators in conjunction with state observers, for example by augmented states (to include parameters) or adaptive observers. 
Parameter estimation alone has been performed by using optimization-based techniques, such as the Levenberg–Marquardt method \cite{Santha-2008} and the Gauss–Newton method \cite{Rama-2011}, as well as parameter identification algorithms such as recursive least-squares \cite{Sun-2015,Zhang-2018b} and instrumental variables \cite{Allafi-2017}. 
Augmented state observers include Kalman filters (KF) \cite{Plett-2004c,Sun-2015,Wang-2017}, particle filters \cite{Restaino-2012} and nonlinear geometric observers \cite{Wang-2015}. 
Adaptive observers combine a state observer for state estimation with another estimator for determining the parameters \cite{Moura-2014,Zhang-2017,Couto-2019,Zheng-2016}. 
For all these observer-based methods, identifiability issues may arise from the state/parameter interplay, and an additional decision system is needed to analyze the evolution of the model parameters as well as evaluate their significance with respect to estimation error.

In contrast with state/parameter estimation, a second approach within model-based methods is residual-based, which involves state observers or parity relations. 
Some approaches deal with state-of-health indication only, where a bank of observers has been used to generate residuals for the fault detection and isolation (FDI) of battery faults like resistance increase \cite{Chen-2014} and over(dis)charge \cite{Sidhu-2015}. 
Others also consider sensor faults and use Kalman filters \cite{Liu-2017,Howey-2017} and sliding-mode observers \cite{Dey-2016}. 
Finally, sensor and/or actuator faults together with battery aging has been addressed via robust observers \cite{Ablay-2014}, Kalman filters \cite{Couto-2018b} and structural analysis \cite{Liu-2016}. 
{However, methods relying exclusively on the innovation sequence are not appropriate to detect parametric changes, because in that case the innovation is not a sufficient statistic\footnote{We say that a statistic $S$ is \textit{sufficient} if the information about the model parameter $\theta$ contained in the sample $\mathcal{Y}$ is concentrated in the statistic $S$. See \cite{Basseville-2003} for more details.}. 
Moreover, methods based on nonlinear parity equations also require the development of an appropriate decision system which might not be easy because the residual pattern will typically depend on the operating conditions.}

In this paper, we depart from the previous works by exploiting a specific type of residual which provides a sufficient statistic for parametric change detection and which lends itself to a sound statistical approach for detection and isolation of small faults via the so-called local approach \cite{Basseville-1993}. 
We investigate how this approach can be applied for early warning of degradation in lithium-ion batteries, which may help to sort batteries in terms of performance at the beginning of life. A reduced-order electrochemical model of the battery is used to obtain a physical interpretation of the parametric changes. We also analyze the correlation between the obtained results for fault detection and isolation and a sensitivity study of the measurements to parametric changes, enabling more insight into the results.

\section{ELECTROCHEMICAL MODELING}

Given its physical basis and therefore the possibility to account for realistic faults, an electrochemical model of the battery cell is considered. 
In this section, we first introduce the reduced-order electrochemical model used to describe the healthy operation of a battery cell, followed by the possible fault modes that we consider in this work.

\subsection{Healthy operation}

An equivalent-hydraulic model (EHM) \cite{Couto-2019} is used to represent the electrochemical behaviour of a battery system. 
In a general sense, this model is mathematically described by the following discrete-time nonlinear dynamical system:
\begin{align}
    \label{xdyn}
    \hspace{-0.2cm}{x}(k+1) & = f(\theta,x(k),u(k)), \\
    \label{yout}
    \hspace{-0.2cm}y(k) & = h(\theta,x(k),u(k)),
\end{align}
where the state vector consists of the state-of-charge and normalized surface concentration $x = [{\rm SOC} \ \ \overline{c}_{ss}]^\top$, 
the input is the applied current $u = I$ and the output is the terminal voltage $y = V$. 
The parameter vector is $\theta = [\varepsilon_s^- \ R_f \ g_s \ n_{Li}]^\top$, which comprises possibly aging-related parameters. 
These parameters are the active material volume fraction $\varepsilon_s$, the film resistance $R_f$, the inverse of the diffusion time constant $g_s$ and the moles of lithium $n_{Li}$. 

Generally, each electrode should be described by an EHM, however here we introduce two common assumptions that reduce the model dynamics to a single EHM. 
First, the positive electrode dynamics are faster than the negative electrode dynamics, which means that the ${\rm SOC}^+$ and  $\overline{c}_{ss}^+$ dynamics are the same as in ${\rm SOC}^+ = \overline{c}_{ss}^+$. 
Secondly, the total moles of lithium in the battery is the sum of the amount of lithium in positive and negative electrodes, i.e.
\begin{equation}
\label{e:balcoef}
\frac{n_{Li}}{A} = c_{s,\mathrm{max}}^+ L^+ \varepsilon_s^+ {\rm SOC}^+ + c_{s,\mathrm{max}}^- L^- \varepsilon_s^- {\rm SOC}^-, 
\end{equation}
which can be solved for ${\rm SOC}^+$ and written as ${\rm SOC}^+ = \theta_1 \rho {\rm SOC}^{\bi{-}} + \theta_4 \sigma$ with $\rho = -c_{s,\mathrm{max}}^- L^- / (c_{s,\mathrm{max}}^+ L^+ \varepsilon_s^+)$ 
and $\sigma = 1/(c_{s,\mathrm{max}}^+ L^+ \varepsilon_s^+ A)$, where $\theta_j$ denotes the $j$-th component of the parameter vector. 
Under these conditions, only the negative electrode state $x = x^-$ is required. 
The nomenclature for all the electrochemical model parameters is reported in the {Appendix}.

The state function is linear and it takes the form
\begin{equation}
    f(\theta,x(k),u(k)) = A(\theta) x(k) + B(\theta) u(k),
\end{equation}
where state matrices, are given by:
\begin{equation}
\label{statef}
A(\theta) \!=\! \left[\!\begin{array}{cc}
1 &0 \\
T_s \theta_3 a_1 & 1 - T_s \theta_3 a_1
\end{array}\!\right]\!, 
B(\theta) \!=\! \frac{T_s}{\theta_1} \left[\!\begin{array}{c}
b_1 \\
b_2
\end{array}\!\right]\!,
\end{equation}
where $T_s$ is the sampling time and with 
\[
a_1 = \displaystyle \frac{1}
{\beta(1-\beta)}, \ \ 
b_1 = \displaystyle \frac{1}{c_{s,\mathrm{max}}^-} \frac{1}{F  L^-}, \ \ 
b_2 = \frac{b_1}{1-\beta}.
\]

The nonlinear output function takes the form
\begin{equation}
\label{hfun}
\begin{array}{rcl}
    h(\theta,x(k),u(k)) &\!\!\!\!= 
    &\!\!\!\! \eta_s^+(\theta,x(k),u(k)) - \eta_s^-(\theta,x(k),u(k)) \\
    &&\!\!\!\!\!\!\!\!\!\!\!\!\!\!\!\!\!\!\!\!\!\!\!\! + U_s^+(\theta,x(k)) - U_s^-(\theta,x(k)) + (\theta_2/\theta_1) d_1 u(k),
\end{array}
\end{equation}
where the nonlinear functions $\eta_s^\pm$ and $U_s^\pm$ are the surface overpotential and the open-circuit potential, respectively, with superscript $+$ for positive and $-$ for negative electrode, and $d_1 = R^-/(3L^-)$. 
Function $\eta_s^\pm$ is given by
\[\eta_s^\pm(\theta,u,x) = \displaystyle\frac{R_g T_{\rm ref}}{\alpha_{0} F} \sinh^{-1} \left( \frac{\mp R^\pm}{6 \varepsilon_s^\pm L^\pm j^\pm_{n,0}(x_\pm)} u \right),\]
where the exchange current density is
\[
j^{\pm}_{n,0}(x^\pm) = k_{n}^{\pm} c_{s,\mathrm{max}}^{\pm} \sqrt{c_{e}} \sqrt{x^{\pm}(\theta) \left( 1 - x^{\pm}(\theta) \right)}.
\]
Function $U_s^\pm$ is empirical and it depends on the considered electrode chemistry \cite{Chaturvedi-2010}. 

This EHM \eqref{xdyn},\eqref{yout} in healthy mode describes the ideal (desired) electrochemical reactions for a lithium-ion battery. 
However, batteries degrade over time due to undesired reactions and material fatigue, which is accounted for next.

\subsection{Faulty operation}

There are different sources of possible degradation that can take place during battery usage. 
Among the most important ones, we can refer to
$(i)$ capacity fade and $(ii)$ power fade, 
which are the faults considered in this work. {Notice that the battery cycle life can be very different from battery to battery, even if they come from the same manufacturing batch \cite{Baumhofer-2014}. Early detection of small internal faults can help to separate the batteries with the best performance from the rest, which would narrow the battery cycle life variations.}

Some electrochemical mechanisms that are responsible for these aging phenomena have been identified, while others are difficult to pinpoint. In the following, we introduce side reactions, as they are the most relevant and widely-known mechanism that degrades Li-ion batteries, and then we explain how we account for other, more uncertain, mechanisms.

\subsubsection{Side reactions} 
In Li-ion batteries, not all the available lithium is effectively used to charge/discharge the battery but part of it is irreversibly consumed in side reactions. 
These reactions compete with the battery desired intercalation reactions, which can be modelled with Kirchhoff's law as
\begin{align}
\label{kir1}
\hspace{-0.2cm} z(k) &= u(k) + d(k) \\
\label{kir2}
\hspace{-0.2cm} 0 &= g(\theta,x(k),u(k),d(k))
\end{align}
where $z(t)$ is the total current flowing through the battery terminals, and $u(t)$ and $d(t)$ are the internal main intercalation reaction and undesired side reaction currents, respectively. 
{The nonlinear function in \eqref{kir2} takes the form
\begin{equation}
\label{hfunsr}
\begin{array}{rcl}
    \!\!g(\theta,x(k),u(k),d(k)) &\!\!\!\!\!= 
    &\!\!\!\!\! U_s^-(\theta,x(k)) \\ [1mm]
    &&\hspace{-2.0cm} + \eta_s^-\!(\theta,x(k),u(k)) - U_{sr} -\eta_{sr}\!(d(k)) ,
\end{array}
\end{equation}
which represents the side reaction taking place at the \mbox{negative} electrode surface \cite{Ramadass-2004,Reniers-2019}, where $U_{sr}$ is a constant that characterizes the occurrence of the side reaction and $\eta_{sr}(d(k))$ is given by
\[
\eta_{sr}(d) = \displaystyle - \frac{R_g T_{\rm ref}}{\alpha_{0} F} \ln \left(- \frac{R^-}{3 
L^- j_{sr,0}} d  \right).
\]}

Under side reaction conditions, the nonlinear output function $h(\theta,x(k),u(k))$ in \eqref{hfun} is now given by
\begin{equation}
\label{hfun2}
\begin{array}{rcl}
    \!\!\overline{h}(\theta,x(k),u(k),z(k)) &\!\!\!\!\!= 
    &\!\!\!\!\! \eta_s^+\!(\theta,x(k),z(k)) \!-\! \eta_s^-\!(\theta,x(k),u(k)) \\
    &&\hspace{-3.5cm} + U_s^+(\theta,x(k)) - U_s^-(\theta,x(k)) + (\theta_2/\theta_1) d_1 z(k),
\end{array}
\end{equation}
which associates the total current $z(k)$ to the positive electrode and only the main intercalation reaction current $u(k)$ to the negative electrode. 
Notice that only the negative electrode suffers from the side reaction, since part of the total current $z(k)$ is lost through $d(k)$ while the positive electrode exploits all the current, i.e. $z(k)$ effectively reflects the main intercalation reaction in the positive electrode. 

The amount of capacity lost due to the side reaction is
\[
Q_{{\rm loss}}(k+1) = Q_{{\rm loss}}(k) -\frac{A}{3600} d(k),
\]
which decreases the lithium inventory $n_{Li}$ according to
\[
n_{Li,t}(k) = n_{Li} - \frac{3600}{F} Q_{\rm loss}(k)
\]
affecting the cell balance of lithium (note that $n_{Li,t}(k)$ substitutes for $\theta_4 = n_{Li}$ when simulating the side reaction). The film resistance growth due to lithium consumption is ignored here.

\subsubsection{Parameter variations} 
Besides side reactions, other degradation phenomena are likely to occur in Li-ion batteries, such as loss of active material, electrode morphological changes, electrolyte decomposition, and others \cite{Onori-2020,Reniers-2019}. 
Even if some efforts have been devoted to identify these mechanisms and derive models to describe them, most of these models are empirical in nature \cite{Reniers-2019}, do not account for each source of degradation and do not fully characterize their complex intertwined behaviour. 
Therefore, instead of considering explicit models for other degradation sources {such as in point 1) above}, we evaluate the performance of the proposed FDI system through specific parametric changes affecting $\theta = [\varepsilon_s \ R_f \ g_s \ n_{Li}]^\top$. 
These parameter variations physically reflect common sources of battery deterioration, like loss of active material $\varepsilon_s$, impedance increase $R_f$, sluggish diffusion $g_s$ and loss of lithium inventory $n_{Li}$. 

\section{SENSITIVITY ANALYSIS}

{In this section, we present a sensitivity analysis \cite{Villa-2016} in order to  study the identifiability of the battery model.} 
This study aims at evaluating the difficulty of estimating a given parameter from real data , i.e.\ it investigates how informative is the data through the sensitivity of measured signals with respect to parameters, as well as the linear correlations among parameters. 
On the one hand, if a model output is not sensitive to a given parameter change, then the parameter cannot be identified. 
On the other hand, if two sensitivity functions are linearly dependent, then the parameters are correlated and it is difficult to identify them individually.

Let us consider the discrete-time system \eqref{xdyn},\eqref{yout}. 
The discrete-time sensitivity variables of the state $x \in \mathbb{R}^2$ and output $y \in \mathbb{R}$ with respect to the parameters $\theta \in \mathbb{R}^4$ are given by
\begin{equation}
\label{sensdt}
s^{x}(k) = \displaystyle \frac{\partial x(k)}{\partial \theta} {\in \mathbb{R}^{2\times4}}, \ \
s^{y}(k) = \displaystyle \frac{\partial y(k)}{\partial \theta} {\in \mathbb{R}^{1\times4}}.
\end{equation}

These sensitivity variables evolve according to the sensitivity system associated with the model \eqref{xdyn},\eqref{yout}, which takes the form
\begin{align}
    \label{xdyns}
    \hspace{-0.3cm}s^{x}\hspace{-0.03cm}(\hspace{-0.03cm}k\!+\!1\hspace{-0.03cm}) &\! =\! \nabla_x f(\theta,x(\hspace{-0.03cm}k\hspace{-0.03cm}),u(\hspace{-0.03cm}k\hspace{-0.03cm})) s^{x}\hspace{-0.03cm}(\hspace{-0.03cm}k\hspace{-0.03cm}) \!+\! \frac{\partial}{\partial \theta} f(\theta,x(\hspace{-0.03cm}k\hspace{-0.03cm}),u(\hspace{-0.03cm}k\hspace{-0.03cm})), \\
    \label{youts}
    \hspace{-0.3cm}s^{y}\hspace{-0.03cm}(\hspace{-0.03cm}k\hspace{-0.03cm}) &\! =\! \nabla_x h(\theta,x(\hspace{-0.03cm}k\hspace{-0.03cm}),u(\hspace{-0.03cm}k\hspace{-0.03cm})) s^{x}\hspace{-0.03cm}(\hspace{-0.03cm}k\hspace{-0.03cm}) \!+\! \frac{\partial}{\partial \theta}  h(\theta,x(\hspace{-0.03cm}k\hspace{-0.03cm}),u(\hspace{-0.03cm}k\hspace{-0.03cm})).
\end{align}

The sensitivities in \eqref{sensdt} can be used to build the discrete-time sensitivity matrix \cite{Villa-2016} as
\begin{equation}
\label{dsm}
S^y = [{s}^{y}_1 \ {s}^{y}_2 \ \cdots \ {s}^{y}_{n_\theta}]
\end{equation}
where ${s}^{y}_{j} \in \mathbb{R}^{N}$ is the column vector formed by the sensitivity of the model output with respect to the $j$-th parameter for data of size $N$. 
Matrix $S^y$ 
is analyzed thanks to the following decomposition $(S^y)^\top S^y = D^\top C D$ \cite{Lund-2008}, where matrices $C$ and $D$ are given by
\begin{equation}
    \label{Dmatrix}
    D = {\rm diag}\left(\|s^y_1\|, \|s^y_2\|,\ldots,\|s^y_{n_\theta}\|\right),
\end{equation}
\begin{equation}
\label{Cmatrix}
    C = \left[ \begin{array}{ccccc}
        1 	&c_{12} &c_{13} &\cdots & c_{1,n_{\theta}} \\
        c_{21}  &1	    &c_{23} & & c_{2,n_{\theta}}\\
        c_{31}  &c_{32} &\ddots  & &\vdots\\
        \vdots  &             &            &1 & c_{n_{\theta-1},n_\theta} \\
        c_{n_\theta,1}   &c_{n_\theta,2}   &\cdots &c_{n_\theta,n_{\theta-1}} & 1
    \end{array}\right].
\end{equation}
In matrix $C$, $c_{ji} = \displaystyle \frac{\langle s^y_j,s^y_i \rangle}{\|s^y_j\| \|s^y_i\|}$ for $j \neq i$ with $i = 1,\ldots,n_\theta$ for each $j = 1,\ldots,n_\theta$, 
$\| \cdot \|$ is the Euclidian norm and $\langle \cdot,\cdot \rangle$ is the inner product \cite{Lund-2008}. 
From this matrix decomposition we can infer linear dependencies among parameters. 
If $\vert c_{ji} \vert$ is close to $1$, then parameters $\theta_j$ and $\theta_i$ are strongly linearly dependent. 
Conversely, values of $c_{ji}$ close to zero imply orthogonality \cite{Moura-2014}. 

\section{OBSERVER-BASED RESIDUALS}

We now proceed to design a state observer for the discrete-time EHM, which is then exploited for FDI, characterized by changes in degradation-related parameters. 
Consider a system similar to \eqref{xdyn},\eqref{yout} but as a stochastic system, i.e.\ including noise terms $w$ and $v$ in the state and output equations, respectively, such as in
\begin{align}
    \label{xdyn0}
    \hspace{-0.2cm}x_s(k+1) & = f(\theta, x_s(k),u(k)) +w(k) \\
    \label{yout0}
    \hspace{-0.2cm}y_s(k) & = h(\theta, x_s(k),u(k)) +v(k),
\end{align}
where the noise sequences abide by 
$w(k) \sim \mathcal{N}(0,Q_x)$ and $v(k) \sim \mathcal{N}(0,R_x)$, with 
$\mathbb{E} [ w(k) v(l)^\top ] = 0$, where
$\mathbb{E}(\cdot)$ is the expectation operator.

Due to the system nonlinearities, an unscented Kalman filter (UKF) is designed, which is reported in Table \ref{t:ukf} in the {Appendix}. 
The UKF relies on the assumption that the distribution of the state vector and the observations are Gaussian. 
To present the algorithm, let us define the state as the concatenation of the original state and the process and measurement noise as
\[ x_s^a(k) = [x_s(k)^\top \ w(k)^\top \ v(k)^\top]^\top \in \mathbb{R}^L, \]
where $L = 2 n_x + n_y$ with $n_x$ and $n_y$ as the state and output dimensions, respectively. 
The state distribution is represented by a minimal set of carefully chosen sample points, called sigma points, which are denoted as 
\[ \mathcal{X}^a(k) = [\mathcal{X}(k)^\top \ \mathcal{X}^w(k)^\top \ \mathcal{X}^v(k)^\top]^\top. \]
These points are generated by a deterministic sampling procedure known as unscented transformation. 
The sigma point generation step is in {\eqref{eq:sigmpoints}}, where ${\hat{x}^a_{s}(k-1) \pm \gamma \sqrt{P^a_{x}(k-1)}}$ stands for 
\[ \hat{x}^a_{s}(k-1) \pm \gamma \left( \sqrt{P^a_{x}(k-1)} \right)_l, l = 1,\ldots,L \]
where $(\cdot)_l$ is the $l$-th column of the matrix. 
The sigma points completely capture the true mean and covariance of the prior random variable. 
When the sigma points are propagated through the nonlinear system, a posterior mean and covariance that are accurate up to the second order are obtained. 

The UKF is used to generate residuals for the FDI using the local approach. 
Under the assumption that
\begin{enumerate}
    \item[(A1)] the parameter $\theta$ is locally identifiable at the nominal value $\theta = \theta_0$ \cite{Zhang-1998a},
\end{enumerate}
we can define a residual related to 
$\theta$ as \cite{Zhang-1998a}
\begin{equation}
\label{Hres}
H(\theta,y_s(k), \hat{y}_s(k)) = s^{y}(k)^\top r(k),
\end{equation}
which minimizes the square of the output error. 
This residual is composed of two parts. 
The first one is the sensitivity of the model output $y$ with respect to the parameter vector $\theta$ given by $s^{y}$ in \eqref{sensdt} but evaluated at $\hat{y}_s$. 
The output sensitivity $s^{{y}}(k)$ can be computed by resorting to the sensitivity system \eqref{xdyns},\eqref{youts}, where the estimates of the state $\hat{x}_s$ coming from the UKF are substituted for $x(k)$. 
The second part is the innovation sequence defined as
\begin{equation}
\label{inno}
r(k) = y_s(k) - \hat{y}_s(k).
\end{equation}

\section{FDI DESIGN BASED ON LOCAL APPROACH}

In this section, we consecutively develop a fault detection and fault isolation strategy based on the local approach \cite{Zhang-1998b} by exploiting the residual $H(\theta,y_s(k), \hat{y}_s(k))$ defined in \eqref{Hres}.

Even in simple cases with linear models, the distribution of {$H(\theta,y_s(k), \hat{y}_s(k))$} is often unknown. 
This issue can be bypassed by approximating this distribution asymptotically. 
One relevant approximation consists of assuming a small change in the parameter, which is known as the local approach \cite{Basseville-2003}. 
Following this approach, for residual $H(\theta,y_s(k), \hat{y}_s(k))$ to be a valid primary residual for monitoring parameter vector $\theta$, the following assumptions need to hold \cite{Zhang-1998b}:
\begin{enumerate}
    \item[(A2)] it is a function differentiable in $\theta$;
    \item[(A3)] it satisfies the following conditions:
\begin{align}
    \label{c1}
    \hspace{-0.2cm} \mathbb{E}_\theta \ H(\theta,y_s(k), \hat{y}_s(k)) & = 0 \ \ {\rm if} \ \ \theta = \theta_0, \\
    \label{c2}
    \hspace{-0.2cm} \mathbb{E}_\theta \ H(\theta,y_s(k), \hat{y}_s(k)) & \neq 0 \ \ {\rm if} \ \ \theta \in v(\theta_0) \backslash \theta_0,
\end{align}
where $\mathbb{E}_\theta$ is the expectation when the system parameter value is $\theta$ and $v(\theta_0)$ is a neighborhood of the nominal parameter value $\theta_0$.
\end{enumerate}

This primary residual is then normalized as
\begin{equation}
\label{normres}
\zeta_N(\theta) = \frac{1}{\sqrt{N}}\sum_{k=1}^N H(\theta,y_s(k), \hat{y}_s(k)).
\end{equation}

Assume further that \cite{Zhang-1998b}
\begin{enumerate}
    \item[(A4)] the sensitivity matrix\footnote{This sensitivity matrix $M$ is for the primary residual-parameter relationship and should not be confused with the state-parameter sensitivity matrix $S^y$ in \eqref{dsm}.}
        \begin{equation}
        \label{sensmat}
        M(\theta_0) = \mathbb{E}_{\theta_0} \frac{\partial}{\partial \theta} H(\theta,y_s(k), \hat{y}_s(k))
        \end{equation}
    exists and is full rank;
    \item[(A5)] the limiting covariance matrix
        \begin{equation}
        \label{resvar}
        \Sigma(\theta_0) = \lim_{N\rightarrow \infty} \mathbb{E}_{\theta_0} \zeta_N(\theta_0) \zeta_N^\top(\theta_0).
        \end{equation}
    exists and is positive definite.
\end{enumerate}

\noindent Then, the normalized residual has remarkable asymptotic statistical properties as stated below.

\subsection{Fault detection}

To detect small changes in the system parameters $\theta$, the local approach is used to discern between the following two local hypotheses:
\begin{align}
    \label{h0}
    \hspace{-0.2cm} {\rm H}_0 : \ \ \theta & = \theta_0, \\
    \label{h1}
    \hspace{-0.2cm} {\rm H}_1 : \ \ \theta & = \theta_0 + \frac{\eta}{\sqrt{N}},
\end{align}
where $\eta$ is an unknown but constant vector and $N$ is the sample size of the data.

The normalized residual $\zeta_N(\theta)$ defined in \eqref{normres} can be proven to be asymptotically Gaussian distributed under both hypotheses in \eqref{h0},\eqref{h1} \cite{Zhang-1998a}, 
i.e.\ the following central limit theorem holds when $N \rightarrow \infty$ \cite{Basseville-2003,Benveniste-1990},
\begin{equation}
\label{hyponormres}
\zeta_N(\theta_0) \rightarrow \left\{ 
\begin{array}{rcl}
\mathcal{N}(0,\Sigma(\theta_0)) &{\rm under} &{\rm H}_0, \\
\mathcal{N}(-M(\theta_0)\eta,\Sigma(\theta_0)) &{\rm under} &{\rm H}_1 ,
\end{array}
\right.
\end{equation}
where the matrices $M(\theta_0)$ and $\Sigma(\theta_0)$ are defined in \eqref{sensmat} and \eqref{resvar}, respectively.
Basically, the local approach performs a sensitivity analysis of the residual with respect to a fault, relative to the residual variance.

The asymptotic Gaussianity of the primary residual in \eqref{hyponormres} opens the door to design asymptotically optimum tests between composite hypothesis \cite{Basseville-2003}. For instance, the decision between ${\rm H}_0$ and ${\rm H}_1$ in \eqref{h0},\eqref{h1} can be achieved through the optimum test statistics
\begin{equation}
\chi^2 = \zeta^\top_N \Sigma^{-1} M (M^\top \Sigma^{-1} M)^{-1} M^\top \Sigma^{-1} \zeta_N
\end{equation}
with $\Sigma = \Sigma(\theta_0)$, $M = M(\theta_0)$ and $\zeta_N = \zeta_N(\theta_0)$, 
which is asymptotically $\chi^2$-distributed as $N \rightarrow \infty$, with $n_\theta$ degrees of freedom. 
The limiting $\chi^2$-distribution is central under ${\rm H}_0$ and has a $\eta^\top M^\top \Sigma^{-1} M \eta$ non-centrality parameter under ${\rm H_1}$.

\subsection{Fault isolation}

The isolation of a given fault is performed once the global $\chi^2$-test has detected it. 
Let us assume that $\zeta_N \sim \mathcal{N}(M \eta, \Sigma)$ and the number $n_a$ of elements of $\theta$ subject to change is known. 
Fault isolation is achieved by testing 
among the $n_a$-size subvector of $\eta$ which one is non zero, 
c.f. if detection has taken place there should be one non zero situation. 
To this end, partition $\eta$ as 
\begin{equation}
\eta = \left[ \eta_a^\top \ \ \eta_b^\top \right]^\top
\end{equation}
with $\eta_a \in \mathbb{R}^{n_a}$ and $\eta_b \in \mathbb{R}^{n_b}$. 
Also assume that $\eta_a$ is the sub-vector to be tested. 
Then we use the minmax test for fault isolation (FI), which consists in viewing the parameters in $\eta_b$ as nuisances and statistically rejecting them. 
In this method, the nuisance parameters $\eta_b$ are replaced by their least favorable value, i.e.\ the value that minimizes the power of the test, which is equivalent to the likelihood ratio \cite{Basseville-2003}
\begin{equation}
\label{LL}
\chi^{2*}_a = 2 \ln \frac{\max_{\eta_a,\eta_b} p_{\eta_a,\eta_b}(\zeta)}{\max_{\eta_b} p_{0,\eta_b}(\zeta)}.
\end{equation}
Let
\begin{equation}
F = M^\top \Sigma^{-1} M \nonumber
\end{equation}
and partition it as
\begin{equation}
\nonumber
F = \left[ \begin{array}{cc}
F_{aa} &F_{ab} \\
F_{ba} &F_{bb}
\end{array} \right] = 
\left[ \begin{array}{cc}
M_a^\top \Sigma^{-1} M_a &M_a^\top \Sigma^{-1} M_b \\
M_b^\top \Sigma^{-1} M_a &M_b^\top \Sigma^{-1} M_b
\end{array} \right],
\end{equation}
where $M=[M_a \ M_b]$ and $M_a$ is made of the $n_a$ columns associated to $\eta_a$. 
The residual of the linear regression of $\tilde{\zeta}_a = M_a^\top \Sigma^{-1} \zeta_N$ with respect to $\tilde{\zeta}_b = M_b^\top \Sigma^{-1} \zeta_N$ is
\begin{equation}
\nonumber
\zeta_a^* =\tilde{\zeta}_a - F_{ab} F_{bb}^{-1} \tilde{\zeta}_b.
\end{equation}
Then, the minmax test in \eqref{LL} can be shown to be
\begin{equation}
\chi^{2*}_a = \zeta_a^{*\top} F_a^{*-1} \zeta_a^*
\end{equation}
where $F_a^* = F_{aa} - F_{ab} F_{bb}^{-1} F_{ba}$ is the covariance matrix of $\zeta_a^*$. 
$\chi_a^{2*}$ is a $\chi^2$-test with $n_a$ degrees of freedom.

\section{SIMULATION RESULTS} \label{simres}

In this section, we present the results obtained when considering a simulated Li-ion battery with graphite/LCO chemistry and whose model parameters are publicly available \cite{Newman-2008}. 
The sampling time was set to $T_s = 1$ s. 
The identifiability properties of the EHM are firstly discussed under a drive cycle (DC) discharge/charge scenario. 
Then, we present the fault detection (FD) and fault isolation (FI) results obtained when resorting to the asymptotic local approach.

\subsection{Identifiability analysis}

\begin{figure}[!htb]
	\centering
	 \includegraphics[scale=0.465,trim={1cm 2cm 7.0cm 3.5cm},clip]{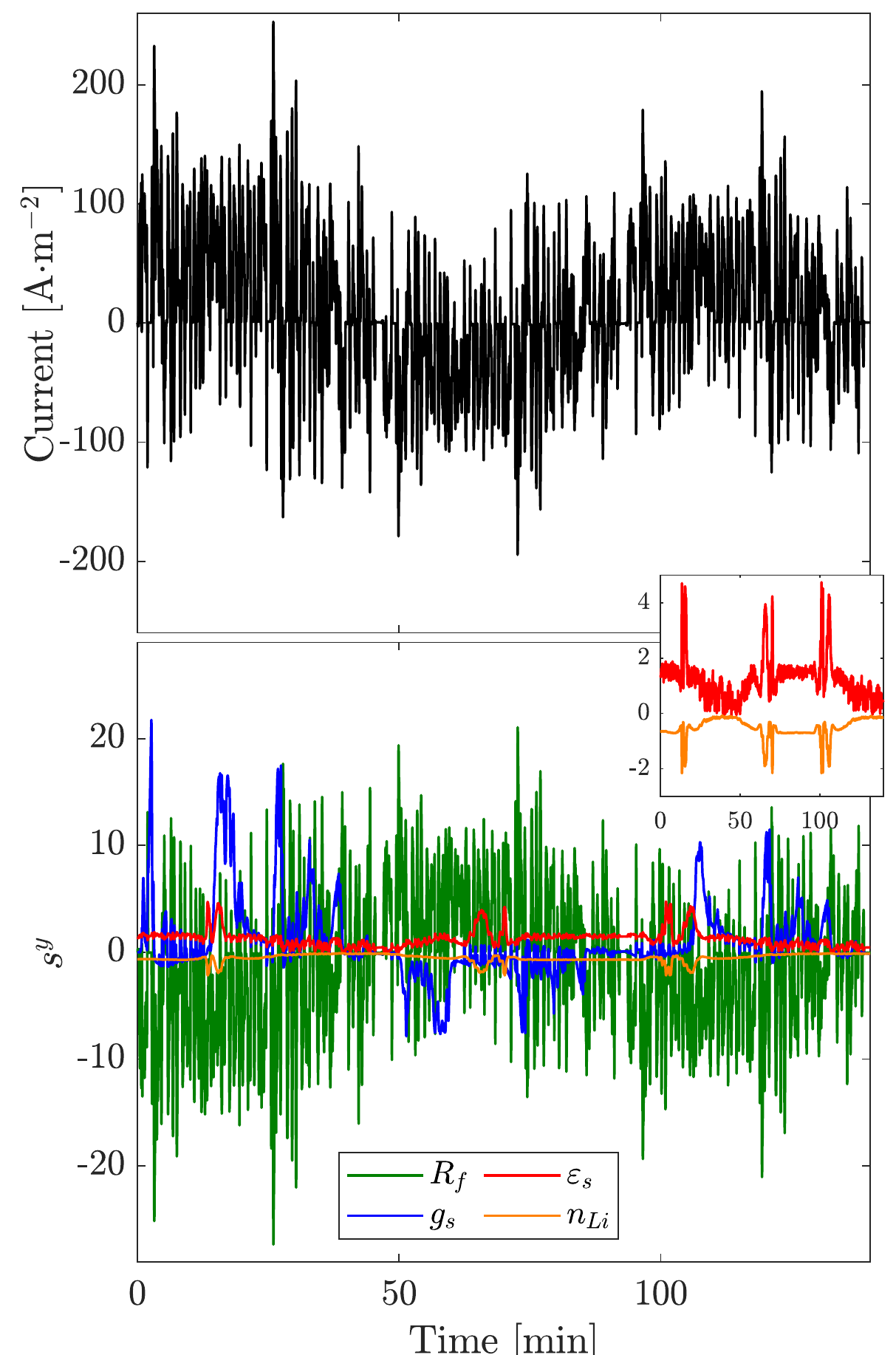}
	 \put (-163,240) {a)}
     \put (-163,125) {b)}
	\vspace{-0.2cm}
    \caption{Relative sensitivity analysis of the EHM with respect to the considered parameter vector for a drive cycle battery discharge/charge. 
	a) Input current profile and b) trajectories of the output sensitivity variables with respect to each parameter.}
	\label{sens}
	\vspace{-0.6cm}
\end{figure}

First, we verified the local structural identifiability of our nonlinear system \eqref{xdyn},\eqref{yout} (assumption A1) through differential geometry \cite{Villa-2019}. 
This check is not reported here due to lack of space. 
Next, we carried out a sensitivity analysis, whose results are shown in Fig.\ \ref{sens}. 
The nonlinear EHM \eqref{xdyn},\eqref{yout} alongside the sensitivity system \eqref{xdyns},\eqref{youts} with sensitivities \eqref{sensdt} were simulated {under a drive cycle load. 
This load consisted of battery discharge/charge (Fig.\ \ref{sens}a) using a UDDS (urban dynamometer driving schedule) profile}, which was scaled (maximum current of $10 C$, $1.8 C$ average) to cover from $0.97$ to $0.25$ SOC of the battery during the first discharge. 
After the first drive cycle discharge, the current profile was inverted for charging and two consecutive charge/discharge cycles were performed. This was done to guarantee a long (more than $5 \times 10^3$ data samples, 80 min) persistent excitation.
Figure \ref{sens}b depicts the trajectories of the output relative sensitivity variables with respect to each considered parameter $\theta \!=\! [\varepsilon_s \ R_f \ g_s \ n_{Li}]^\top\!$. 
From the figure it follows that the sensitivity of $R_f$ is clearly the largest one, followed by the one of $g_s$ with the largest sensitivity levels during a few intervals. 
Then, $\varepsilon_s$ sensitivity follows and $n_{Li}$ is the least sensitive parameter (figure inset). 
This sensitivity ranking is confirmed by matrix $D\!=\!{\rm diag}(134.054, 568.042, 343.068, 59.173)$ in \eqref{Dmatrix}. 

Next, linear dependencies among parameters are assessed by computing the $C$ matrix in \eqref{Cmatrix}, resulting in
\begin{equation}
\label{Cmatrixdcn}
    C = \left[ \begin{array}{cccc}
   1.000   &-0.306    &0.305     &-0.981\\
   -0.306  &  1.000   &-0.315    &0.172\\
    0.305  & -0.315   & 1.000    &-0.203\\
   -0.981  &  0.172   &-0.203    &1.000
    \end{array}\right],
\end{equation}
The most correlated parameters are by far $\varepsilon_s$ and $n_{Li}$ with $c_{14} = -0.981$, followed by $\varepsilon_s, R_f$ and $g_s$ with $|c_{ji}| \approx 0.30$, $j\in\{1,2\}$ and $i\in[2,3]$ with $i\neq j$. 

\begin{table*}[!htb]
\vspace{0.2cm}
\caption{Average results of the global $\chi^2$-test. 
First row: FD results; remainder rows: FI minmax tests. 
From $\chi^2$-table follows: threshold $13.3$ for FD and $6.6$ for FI with $0.01$ probability of false alarm.}
\vspace{-0.4cm}
\label{t:fdidc}
\begin{center}
\begin{tabular}{c c c c c c c c c c}
\hline
 &$\theta_0$ & ${\rm f}_{\varepsilon_s}$ & ${\rm f}_{R_f}$ & ${\rm f}_{g_s}$ & ${\rm f}_{n_{Li}}$ & ${\rm f}_{d(k)}$ & ${\rm f}_{R_f}, 0.2\%$
 & ${\rm f}_{g_s}, 5\%$ & ${\rm f}_{d(k)}$$^{a}$ \\
\hline
&\multicolumn{9}{c}{Fault detection} \\
$\chi^2$ &9.750	&{\bf 39.660}  &{\bf 20.113} &10.310  &{\bf 14.761} &9.673 &{\bf 44.680} &{\bf 16.220} &{\bf 24.972} \\
&\multicolumn{9}{c}{Minmax test} \\								
$\varepsilon_s$	&3.434	&3.456	 &4.041   &3.624  &{\bf 9.144}  &3.198 &3.087  &{\bf 7.953}  &{\bf 10.872} \\
$R_f$		    &3.040	&{\bf 12.631}  &5.494   &2.862  &{\bf 8.305}  &2.617 &{\bf 16.869} &5.901  &{\bf 19.768} \\
$g_s$		    &2.738	&5.429	 &2.997	  &2.128  &2.011  &2.499 &3.797  &{\bf 8.979}  &2.922 \\
$n_{Li}$	    &3.297	&3.375	 &3.937   &3.453  &{\bf 8.831}  &3.097 &2.945  &{\bf 8.017}  &{\bf 10.574} \\
\hline
\end{tabular} \\
$^a$$j_{sr,0} = 3\times10^{-5}$ A$\cdot$m$^{-2}$.
\end{center}
\vspace{-0.6cm}
\end{table*}

\subsection{Fault detection and isolation}

The FDI simulation is based on the EHM \eqref{xdyn},\eqref{yout} as plant model and the drive cycle as input current. 
{Five faulty cases were considered, which are denoted as ${\rm f}_{\theta_j}$ for a fault ${\rm f}$ affecting parameter $\theta_j$ (e.g. ${\rm f}_{\varepsilon_s}$ for $\theta_1=\varepsilon_s$) with  $j=1,\ldots,4$, 
while the fifth fault incorporates the side reaction in the model equations, i.e.\ side reaction current $d(k) \neq 0$, and it is denoted as ${\rm f}_{d(k)}$.} 
{Note that the last fault is outside the formalism presented for the local approach, but it is intended to demonstrate how the algorithm performs when faced against more complex types of faults that might affect battery operation.} 
Unless otherwise stated, the model parameters were varied by $\Delta \theta = 0.1\%$ relative error with respect to their nominal values $\theta_0$ whereas the side reaction exchange current density $j_{sr,0} = 1.5\times 10^{-6}$ A$\cdot$m$^{-2}$ \cite{Ramadass-2004}. 
These parameters translate into a maximum capacity fade of 0.07\% and ohmic drop of 0.004\% when a fully charged battery is discharged in simulation at $C/2$. 
The voltage signal was corrupted with a white noise sequence of variance $R = 10$~mV$^2$ in all cases.

The UKF tuning parameters were selected to be ${\alpha = 0.1}$, $\beta = 2$ and $\kappa = 3 - L$ \cite{Vdm-2002}. The state vector was initialized with 
$-5\%$ error. 
The measurement noise variance was set to its actual value, whereas the process noise variance was set to $Q = 10^{-8} I_2$ and the initial error covariance matrix was $P_{x,0} = 10^{-3} I_2$, which provided good convergence of the filter in simulation.

For each simulation of the FDI system, the sample size was $N = 8400$ from which we discarded the first 200 samples 
to avoid transient effects. 
{The implementation of the algorithm is done in batch, i.e. a given time series of data is processed at a time, and it could be performed periodically for FDI.} 
For each considered parameter value, either nominal or changed, 100 simulations were carried out with different measurement noise realizations. 
The estimation of matrix $\Sigma(\theta_0)$ in \eqref{resvar} from available data is not a trivial problem as pointed out in \cite{Zhang-1998a,Basseville-2003}. 
We computed it here as \cite{Zhang-1998a}
\begin{equation}
\label{Sapprox}
\Sigma \approx \frac{1}{N} \sum_{k=1}^{N} H_k H_k^\top + \sum_{i=1}^{n_i} \frac{1}{N-i} \sum_{k=1}^{N-i} \left( H_k H_{k+i}^\top + H_{k+i} H_k^\top \right). 
\end{equation}
We set 
$n_i = 12$, which ensures that $\Sigma$ is positive definite. 
Matrix $M$ in \eqref{sensmat} is estimated by \cite{Zhang-1998a}
\begin{equation}
\label{Mapprox}
M \approx \frac{1}{N} \sum_{k=1}^N \left[  - s^y(k)^\top s^y(k) + 
r(k) s^{yy}(k) \right]_{\theta = \theta_0}.
\end{equation}
where $s^{yy}$ denotes second-order sensitivities of the model output $y$ with respect to the parameter vector $\theta$. 
The sensitivity equations were computed with CasADI \cite{Andersson-2019}, which is automatic differentiation software able to calculate first and second-order derivatives efficiently.

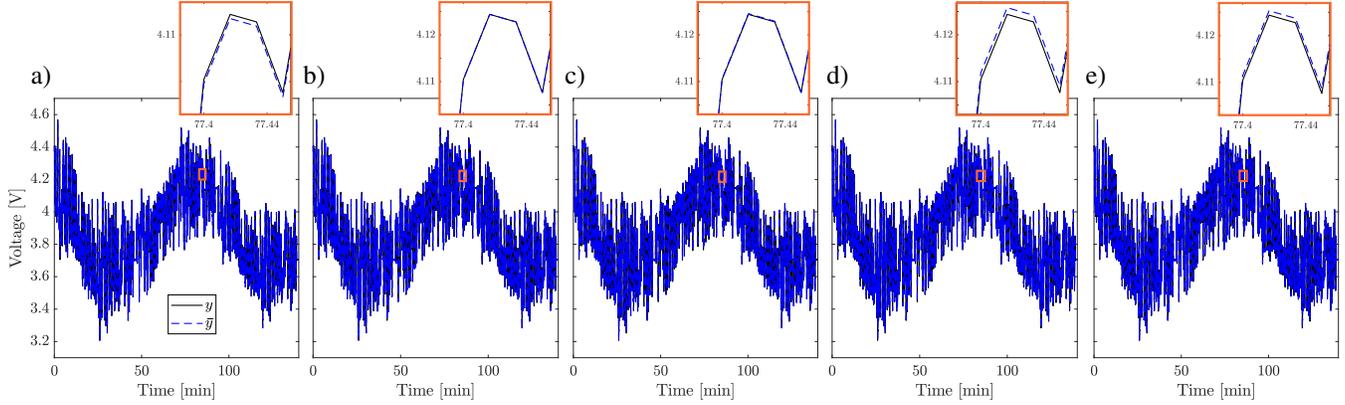
\begin{figure*}[!hbt]
	\centering
	\vspace{0.3cm}
	\includegraphics[scale=0.4,trim={0.5cm 6cm 0.5cm 1.5cm},clip]{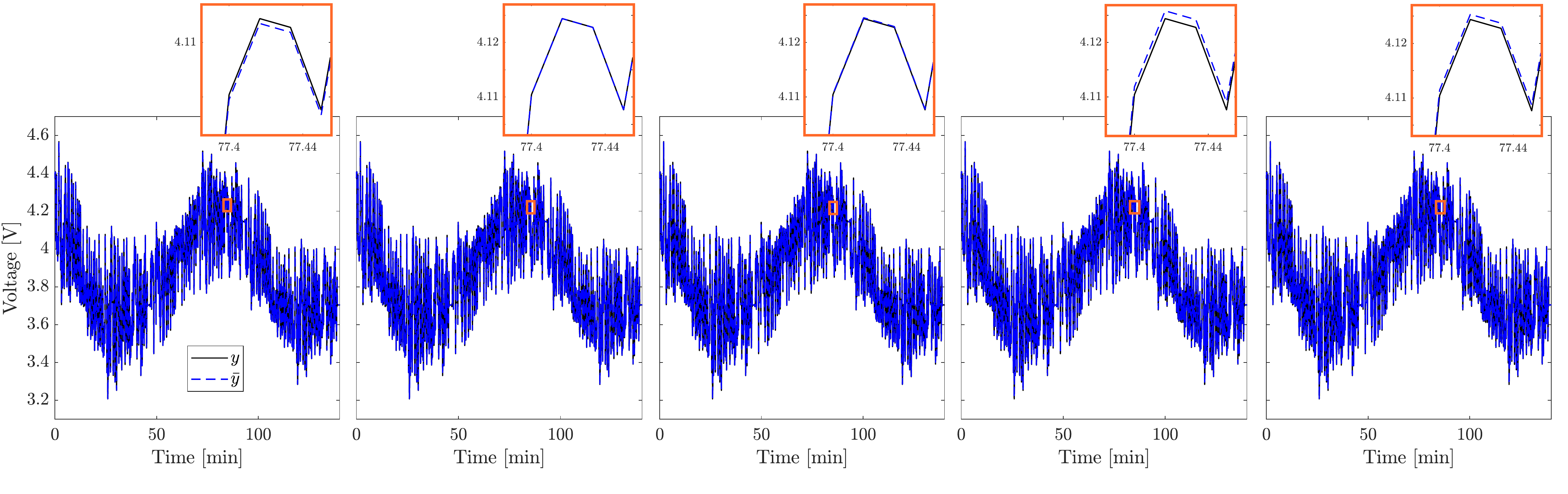}
	 \put (-500,128) {a)}
     \put (-397,128) {b)}
     \put (-298,128) {c)}
     \put (-199,128) {d)}
     \put (-101,128) {e)}
     \vspace{-0.6cm}
	\caption{Voltage output responses under drive cycle. Both healthy $y$ and faulty $\bar{y}$ modes are shown. 
	Different faulty behaviours are depicted, namely parameter variations of a) $\Delta\theta = 0.1\%$ in $\varepsilon_s$ and b) $R_f$, c) $\Delta\theta = 5\%$ in $g_s$, d) $\Delta\theta = 0.1\%$ in $n_{Li}$ and e) side reaction of $j_{sr,0} = 3\times 10^{-5}$ A$\cdot$m$^{-2}$.}
	\label{f:dcres}
	\vspace{-0.4cm}
\end{figure*}

Table \ref{t:fdidc} shows the FD results for the $\chi^2$-test and the FI results for the minmax test under nominal (first column entry) and slightly changed ($\Delta\theta=0.1\%$ in second to sixth column entries) parameter values. 
{That level of parameter variation translates into overall faults of ${\rm f} = [-600 \ 10 \ -3 \ -2000] \times 10^{-6}$ where ${\rm f} = [{\rm f}_{\varepsilon_s} \ {\rm f}_{R_f} \ {\rm f}_{g_s} \ {\rm f}_{n_{Li}} ]$.} 
From these results it follows that it is possible to detect faults affecting $\varepsilon_s, R_f$ and $n_{Li}$. 
{The largest $\chi^2$ value is obtained for $\varepsilon_s$ instead of the most sensitive parameter $R_f$, which is due to the 60 times larger fault in absolute value for the former parameter. 
Similarly, despite the large and small sensitivities of $g_s$ and $n_{Li}$, respectively, their faults are the smallest and biggest ones which prevents and helps their respective detection. 
No parameter can be correctly isolated, with $\varepsilon_s$ fault ascribed to $R_f$.} 
Moreover, $n_{Li}$ is associated with changes in $n_{Li}$ itself, $\varepsilon_s$ and $R_f$, which can be explained by the low sensitivity of $n_{Li}$ and its high correlation with $\varepsilon_s$, while the influence of $R_f$ seems to be limited given the lower value of its indicator w.r.t. $n_{Li}$ and $\varepsilon_s$. 
In order to test if larger fault magnitudes can be picked up by the algorithm for the remaining undetected and non-isolated faults, we increase $\Delta \theta = 0.2\%$ for $R_f$, $\Delta \theta = 5\%$ for $g_s$ and {$j_{sr,0} = 3\times 10^{-5}$ A$\cdot$m$^{-2}$ for the side reaction}. 
{The results are shown in the last three columns of Table \ref{t:fdidc}. 
{Under these changes, now the $\chi^2$-test is able to detect these faults.} 
While the isolation of $R_f$ is now correct, the indicator of $g_s$ becomes activated for $\varepsilon_s$, $n_{Li}$ and $g_s$ itself with {the latter having a slight edge over the former ones.} 
Finally, the side reaction is ascribed mainly to $R_f$. 
Since the side reaction can be seen as a perturbation to the input current and $R_f$ appears as input coefficient in the model, this allocation is logical.}

To give an idea of how small the considered fault magnitudes are, and how difficult is to detect/isolate them, 
\mbox{Fig.\ \ref{f:dcres}}a-e shows the model output when the system is healthy (solid black curves) and when it is subject to a given fault (dashed blue curves). 
The five faulty scenarios are depicted. 
The worst fault cases in Table \ref{t:fdidc} are considered for each parameter. 
The figure shows that the different faults are imperceptible from the voltage standpoint for the naked eye (overlapping curves), with a capacity fade and an ohmic drop of up to 0.15\% and 0.004\%, respectively. 
Only when a considerable zoom is applied (figure insets) it is possible to see slight changes in the voltage trajectories from healthy to faulty modes.

\vspace{-0.1cm}
\section{CONCLUSIONS}
\vspace{-0.1cm}

A diagnosis scheme aiming at detecting and isolating small faults characterizing lithium-ion battery degradation has been reported. This scheme relies on the asymptotic local approach. First, a reduced-order electrochemical model of the battery under both healthy and faulty conditions was used, and its identifiability properties were verified. Secondly, a state observer was used to generate primary residuals along with the limiting covariance matrix. Finally, the residuals were exploited by $\chi^2$-tests for FDI. 
It was verified that the sensitivity of the parameters is large with an aggressive drive cycle. 
{All the considered parameter changes can be successfully detected, even if some relative faults need to be larger than $0.1\%$ to do so. 
The most sensitive parameters are the volume fraction and moles of lithium, which facilitates the FDI of these faults.} 
On the other hand, the isolation of less sensitive and highly correlated parameters might be erroneous. 
Overall, detection of small degradation rates causing less than $0.15\%$ capacity fade and $0.004\%$ resistance increase can be achieved, {which is well below the accuracy of experimental procedures to measure capacity/resistance.} 
The proposed method might also be particularly useful for the early warning of  
{small battery internal faults that may condition the entire cycle life of a battery, which we will pursue in future work.}






\vspace{-0.1cm}
\section*{{APPENDIX}}
\vspace{-0.1cm}

The nomenclature and UKF algorithm are now reported.



\begin{table}[!ht]
\vspace{0.3cm}
\caption{Nomenclature.}
\vspace{-0.2cm}
\centering
  \begin{tabular}{c c}
  \hline
  Parameter  & Symbol \\ \hline
  Cross-sectional area [m$^{2}$]
             & $A$  \\ 
  Maximum lithium concentrations [mol.m$^{-3}$]
             & $c_{s,{\rm max}}^\pm$  \\ 
  Electrolyte concentrations [mol.m$^{-3}$]
             & $c_e$  \\ 
  Faraday's constant [C$\cdot$mol$^{-1}$]
             & $F$       \\
  Diffusion time constant inverse [s$^{-1}$]
             & $g_s$  \\ 
  Side reaction exchange current density [A.m$^{-2}$] 
             & $j_{sr,0}$   \\ 
  Reaction rate constant [A.m$^{2.5}$.mol$^{-1.5}$] 
             & $k_n^\pm$   \\ 
  Electrode thickness [m$^{2}$] 
             & $L^\pm$   \\ 
  Particle Radius [m] 
             & $R^\pm$   \\ 
  Film resistance [$\Omega$.m$^2$] 
             & $R_f$      \\
  Universal gas constant [J$\cdot$mol$^{-1}\cdot$K$^{-1}$]
             & $R_g$      \\ 
  Reference temperature [K]
             & $T_{\mathrm{ref}}$  \\
  Apparent transfer coefficient [-]
             & $\alpha_0$        \\
  Particle volume ratio [-]
             & $\beta$        \\
  Active material volume fraction [-]
             & $\varepsilon_s$        \\ \hline
  \end{tabular}
\label{tab:func}
\end{table}

\begin{table}[!ht]
\vspace{0.3cm}
\caption{UKF for the nonlinear model in \eqref{xdyn0},\eqref{yout0}$^\dagger$.}
\vspace{-0.2cm}
\centering

\begin{tabular}{l}
\hline \vspace{-0.2cm}
\\
Initialization: for $k = 0$, set \\
\forceindent $\hat{x}_{s,0} = E [ x_{s,0} ]$, \ \ 
$P_{x,0} = E[ (x_{s,0} - \hat{x}_{s,0}) (x_{s,0} - \hat{x}_{s,0})^\top ]$ \\
\forceindent $\hat{x}_{s,0}^a = E[ x_{s,0}^a ] = [ \hat{x}_{s,0}^\top \ \ 0 \ \ 0 ]^\top$ \\ 
\forceindent $P_{x,0}^a = E[ (x_{s,0}^a - \hat{x}_{s,0}^a)(x_{s,0}^a - \hat{x}_{s,0}^a)^\top ] = 
\mathrm{diag}(P_{x,0}, Q_x, R_x)$ 
\\
Computation: for $k = 1, 2, \ldots$ compute \\

Sigma points: 
\\
\vbox{
\begin{align}
\label{eq:sigmpoints}
\hspace{-0.2cm} \mathcal{X}^a_{k-1} & = 
\left[ \hat{x}^a_{s,k-1} 
\ \ \hat{x}^a_{s,k-1} + \gamma \sqrt{P^a_{x,k-1}} 
\ \ \hat{x}^a_{s,k-1} - \gamma \sqrt{P^a_{x,k-1}} \right] 
\end{align}
} 
\vspace{-0.2cm}  \\
Time-update: \vspace{-0.2cm} \\
\vbox{
\begin{align}
\label{eq:timeup1}
\hspace{-0.2cm} \mathcal{X}_{k|k-1} &= 
f \left( \theta, \mathcal{X}_{k-1},u_{k} \right) + {\mathcal{X}^{w}_{k-1}} \\
\label{eq:timeup2}
\hspace{-0.2cm} \hat{x}^-_{s,k} & = 
\sum_{l=0}^{2L} W_{l}^{(m)}\mathcal{X}_{l,k|k-1} \\
\label{eq:timeup3}
\hspace{-0.2cm} P^-_{x,k} & = 
\sum_{l=0}^{2L} W_{l}^{(c)} \left( \mathcal{X}_{l,k|k-1} - \hat{x}^-_{s,k} \right) \left( \mathcal{X}_{l,k|k-1} - \hat{x}^-_{s,k} \right)^\top
\end{align}
} \vspace{-0.2cm} \\
Measurement-update \vspace{-0.2cm} \\
\vbox{
\begin{align}
\label{eq:timeup4}
\hspace{-0.2cm} \mathcal{Y}_{k|k-1} &= 
h \left(\theta, \mathcal{X}_{k|k-1}, u_{k} \right) + \mathcal{X}^{v}_{k-1} \\
\label{eq:timeup5}
\hspace{-0.2cm} \hat{y}_{s,k} &= 
\sum_{l=0}^{2L} W_{l}^{(m)}\mathcal{Y}_{l,k|k-1} \\
\label{eq:measup1}
\hspace{-0.2cm} P_{y,k} &= 
\sum_{l=0}^{2L} W_{l}^{(c)} \left( \mathcal{Y}_{l,k|k-1} - \hat{y}_{s,k} \right) \left( \mathcal{Y}_{l,k|k-1} - \hat{y}_{s,k} \right)^\top \\
\label{eq:measup2}
\hspace{-0.2cm} P_{xy,k} &= 
\sum_{l=0}^{2L} W_{l}^{(c)} \left( \mathcal{X}^{x}_{l,k|k-1} - \hat{x}^-_{s,k} \right) \left( \mathcal{Y}_{l,k|k-1} - \hat{y}_{s,k} \right)^\top \\
\label{eq:measupmid}
\hspace{-0.2cm}\mathcal{K}_{k} & = 
P_{xy,k} P^{-1}_{y,k}, \\
\label{eq:stateestim}
\hspace{-0.2cm} \hat{x}_{s,k} &= 
\hat{x}^-_{s,k} + \mathcal{K}_{k} \left( y_{s,k} - \hat{y}_{s,k} \right), \\
\label{eq:covestim}
\hspace{-0.2cm} P_{x,k} &= 
P^-_{x,k} - \mathcal{K}_{k} P_{y,k} \mathcal{K}_{k}^\top, 
\end{align}
} \vspace{-0.2cm} \\
Tuning parameters: \vspace{-0.2cm} \\
\vbox{
\begin{equation}
\begin{array}{l}
\gamma = \sqrt{L + \lambda}, \ \ \lambda = \alpha^2(L+\kappa)-L, \\
W^{(m)}_{0} = \frac{\lambda}{L + \lambda}, \ \ W^{(c)}_{0} = \frac{\lambda}{L + \lambda} + 1 - \alpha^2 + \beta, \\
W^{(m)}_{l} = W^{(c)}_{l} = \frac{1}{2 (L + \lambda)}, \ \ l = 1,\ldots,2L, L = 2n_x+n_y
\end{array}
\end{equation}
} \vspace{-0.3cm} \\ 
\hline \\
\vbox{\vspace{-0.4cm}
\begin{flushleft}
$^\dagger$For convenience, the discrete time argument $k$ is written as a subscript in the table.
\end{flushleft}
}
\vspace{-1cm}
\end{tabular}

\label{t:ukf}
\end{table}




\bibliographystyle{unsrt}        
\bibliography{Reference}

\end{document}